\documentclass{PoS}
\usepackage{graphicx}
\usepackage{placeins}
\usepackage[labelfont=bf, labelsep=period]{caption}
\usepackage{subfigure}
\usepackage{wrapfig}
\usepackage{url}
\usepackage{stackengine}
\usepackage{bm}

\title{Calibration LEDs in the IceCube Upgrade D-Egg Modules}

\ShortTitle{Calibration LEDs in D-Egg}

\author{
The IceCube Collaboration\footnote{For collaboration list, see PoS(ICRC2019) 1177.}\\
{\itshape \href{http://icecube.wisc.edu/collaboration/authors/icrc19_icecube}{http://icecube.wisc.edu/collaboration/authors/icrc19\_icecube}}\\
E-mail: \email{ayumi\_kiriki@hepburn.s.chiba-u.ac.jp, aya@hepburn.s.chiba-u.ac.jp}
}

\abstract{The IceCube Upgrade, planned for deployment in the 2022/2023 South Pole Summer, will involve deployment of a greater density of optical modules (vertically spaced $\sim$3\,m).
Improvements in the calibration of optical sensors and an enhanced understanding of the optical properties of the deep glacial ice, due to the more closely-spaced modules, are projected to have a large impact on neutrino reconstruction.
A new optical sensor module called the ``Dual optical sensor in an Ellipsoid glass for Gen2'' (D-Egg), is planned for installation in both the IceCube Upgrade and IceCube-Gen2, and has both an upward and downward facing 8'' high quantum efficiency PMT.

The D-Egg modules will make use of a downward-facing LED calibration system to measure the optical properties of the refrozen drill holes (``hole ice''). 
An inner section of the hole ice contains some impurities, which modify the optical properties of the ice; this area is known as the ``bubble column''. The measurement of this ``hole ice'' is critical both for the upgraded IceCube detector as well as the current generation IceCube, as refrozen ice contributes significant systematic uncertainties to the reconstruction of low energy neutrinos. A simulation was performed, where the size and optical properties of the bubble column were varied. A log likelihood function is constructed from the geometry of the D-Eggs and properties of the hole ice. The minimization recovers best fit values close to the Monte Carlo truth. 
\vspace{4mm}
{\bfseries Corresponding authors:} 
A.~Kiriki$^{1}$,
\speaker{A.~Ishihara}$^{\;2}$\\
{$^{1}$ \itshape Dept. of Physics, Division of Advanced Science and Engineering, Graduate School of Science and Engineering, Chiba University, Chiba 263-
8522, Japan}\\
{$^{2}$ \itshape Dept. of Physics and Institute for Global Prominent Research, Chiba University, Chiba 263-8522, Japan}

}

\FullConference{36th International Cosmic Ray Conference -ICRC2019-\\
		July 24th - August 1st, 2019\\
		Madison, WI, U.S.A.}

\begin{document}

\section{The IceCube Upgrade and hole ice}

\begin{wrapfigure}[13]{r}[2pt]{6.5cm}
\centering
\includegraphics[width=.95\linewidth]{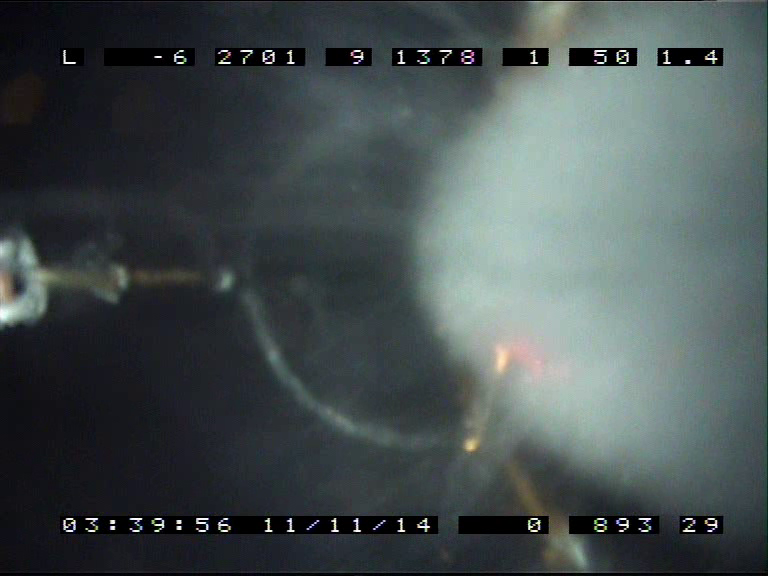}
\caption{Camera image looking downward into the hole ice. The bubble column is found on the center right.}
\label{fig:camera}
\end{wrapfigure}

The IceCube Neutrino Observatory \cite{ref:icecube-jinst} currently consists of 5160 digital optical modules (DOMs) inserted deep into the glacial ice at the South Pole. Each DOM has a diameter of 33\,cm, a single downward-facing 10'' photomultiplier tube (PMT), and detects the Cherenkov light produced by charged particles passing through the ice, the most interesting of which result from neutrino-nuclei interactions. The DOMs were deployed at depths between 1450\,m and 2450\,m across 86 holes of 60\,cm diameter, where the glacial ice had been melted with a hot water drill.
The refrozen ice, called ``hole ice'', is expected to have different optical properties than the undisturbed glacial ice (bulk ice).\par

The hole ice is composed of two regions: a central column containing some amount of impurities, called the bubble column, and a clear outer region (Figure~\ref{fig:camera}).
The bulk ice has $\sim$20\,m long scattering length \cite{ref:bulk}, while the bubble column has a much shorter scattering length. This scattering length depends on the size and distribution of the impurities.
The properties of the hole ice can be characterized by two parameters; scattering in the bubble column and the bubble column size.
Current IceCube DOMs do not have fully vertical light emitting diodes (LEDs), and therefore are not optimal for directly measuring the hole ice. 
Combined with the large spaces between DOMs (vertically 17\,m, horizontally 125\,m), previous measurements of the hole ice properties remain inconclusive \cite{ref:previous_study}.
Lack of understanding regarding the hole ice properties is detrimental to the direction reconstruction for all neutrinos, and is particularly relevant for $\nu_e$, $\nu_{\tau}$, and neutral current interactions in the GeV region \cite{ref:uncertainty}.
The IceCube Upgrade \cite{ref:upgrade} aims to resolve this problem and decrease such systematic uncertainties by improving the understanding of ice properties enabled by new optical modules.
These new optical modules are planned to be deployed at a depth between 2150\,m and 2425\,m with a vertical spacing of 2.7\,m. One type of these new optical modules is called the Dual optical sensor in an Ellipsoid Glass for Gen2 (D-Egg) (Figure~\ref{fig:degg_img}) \cite{ref:degg-icrc2017}. 
The D-Egg modules are also planed to be deployed in the future IceCube-Gen2 experiment, which looks to increase the detection rate of astrophysical neutrinos, which in turn enables more sensitive multi-messenger searches \cite{ref:icecube-gen2}.

\section{D-Egg module in the IceCube Upgrade}
300 D-Eggs are planned for deployment in the 2022/2023 South Pole Summer as part of the IceCube Upgrade.
A D-Egg module is 30\,cm in diameter.
Each D-Egg module contains two 8'' high quantum efficiency PMTs (Hamamatsu Photonics R5912-100-20), one facing upward and the other downward. They are housed in an ellipsoidal glass container and are optically coupled to the glass with optical elastomers.\par

\begin{figure}
  \centering
\subfigure[]{\includegraphics[width=.37\linewidth]{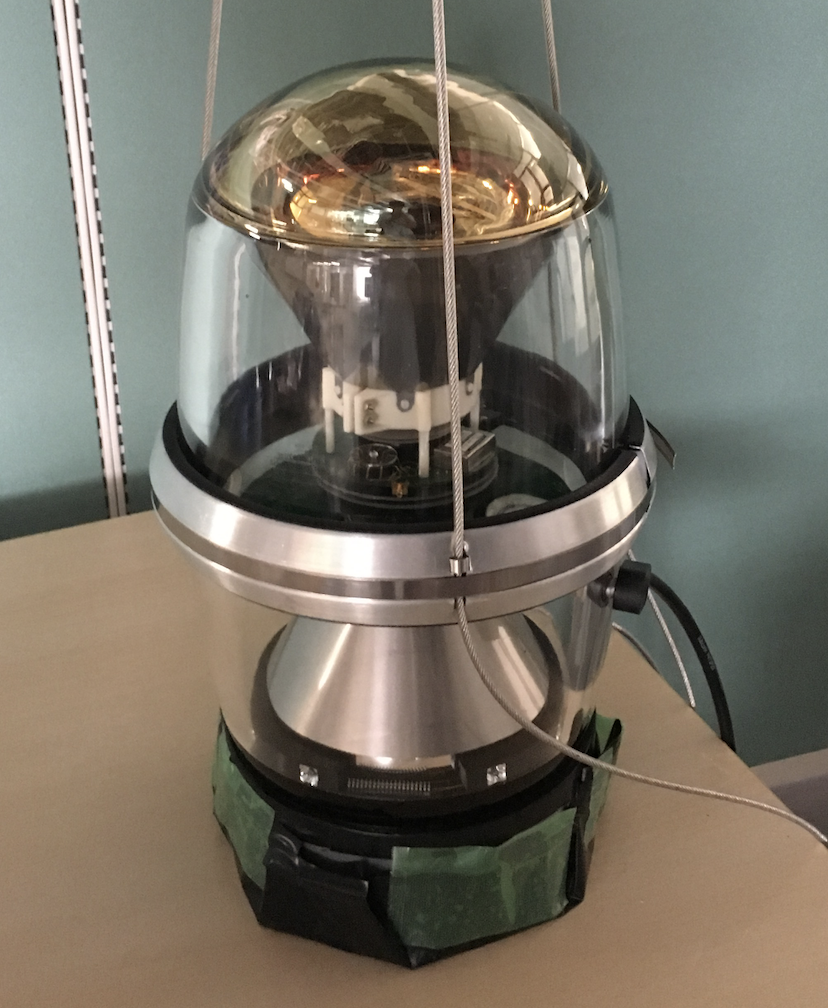}\label{fig:degg_img}}
\hspace{10mm}
\centering
\subfigure[]{\includegraphics[width=.4\linewidth]{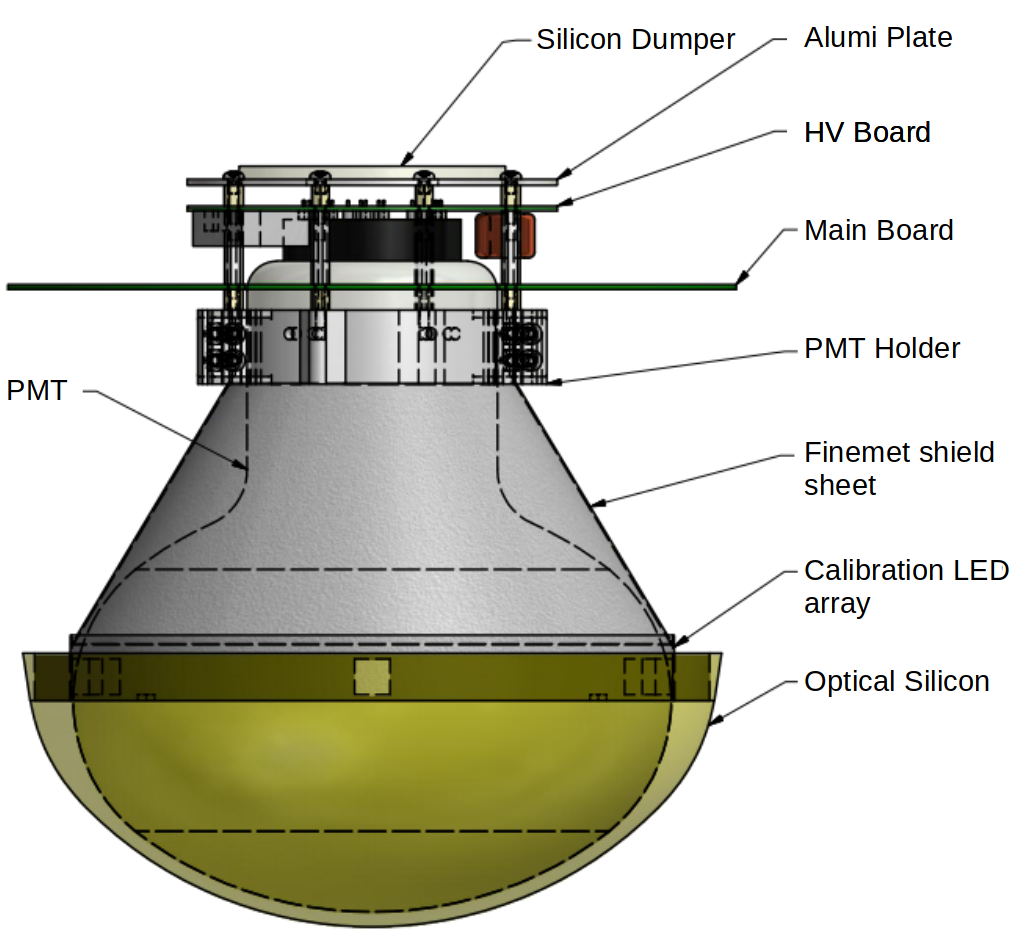}\label{fig:degg_cad}}
\caption{a) Prototype D-Egg. b) Section drawing of the lower structure. The flasher holder is placed on the lower optical elastomer. Eight LEDs shooting horizontally are implanted along the side and four LEDs illuminating vertically at the bottom.}
\end{figure}

The D-Egg module includes 12 LEDs for calibration purposes, eight of which are pointing in the horizontal plane, and among other things, will be used to calibrate the orientation of the optical modules. The other four LEDs are downward-facing and will be used to calibrate the hole ice.
All LEDs will be installed in a special ring called the ``flasher holder'', such that the four vertical LEDs with $90^\circ$ spacing are attached to the lower optical elastomer (Figure~\ref{fig:degg_cad}). 
The light emitted by the downward-facing LEDs will then be detected by the D-Eggs in the same hole.
Because the optical properties of the hole ice largely depend on the size of the bubble column and the position of the D-Egg relative to it, these four LEDs can reduce the overall uncertainty of the hole ice. %ease uncertainties in the relative position of a D-Egg to a bubble column.
Specifically, measurements with D-Eggs will determine the bubble column diameter $D$ and the effective scattering length of the bubble column $\lambda_\mathrm{e}$.
Although, D-Eggs will only be deployed in future IceCube expansions, these results plan to be applied to the entire IceCube data collected over 10 years.

\section{Monte Carlo simulations} \label{monte_carlo_simulation}

\subsection{Photon Propagation}

Photon propagation has been simulated using Geant4 \cite{ref:geant4}.
To study the photon propagation in the bubble column, a large volume of bulk ice, which does not scatter photons, is generated.
The drill hole is then simulated with a fixed diameter of 60\,cm (in which the D-Eggs will be placed), and the bubble column is simulated in the center of the drill hole with a diameter $D$. The clear outer region of the hole ice has similar properties to the bulk ice, so the interface between the two is not significant.
To parameterize the Mie scattering in the bubble column, the Henyey-Greenstein scattering function \cite{ref:HGfunc} is implemented with an average scattering angle of $\left<\cos{\theta}\right>=0.95$.
The effective scattering length ($\lambda_\mathrm{e}$) in the simulated ice column, where $\lambda_\mathrm{s}$ is the geometrical scattering length, is defined as:
\begin{eqnarray}
\lambda_\mathrm{e}=\lambda_\mathrm{s}/\left(1-\left<\cos\theta\right>\right)~.
\end{eqnarray}

\subsection{Simulation Geometry}%or some other name

\begin{figure}
\begin{tabular}{c@{\hspace{1cm}}c}
 \begin{tabular}{c}
  \subfigure[]{\includegraphics[width=.46\linewidth]{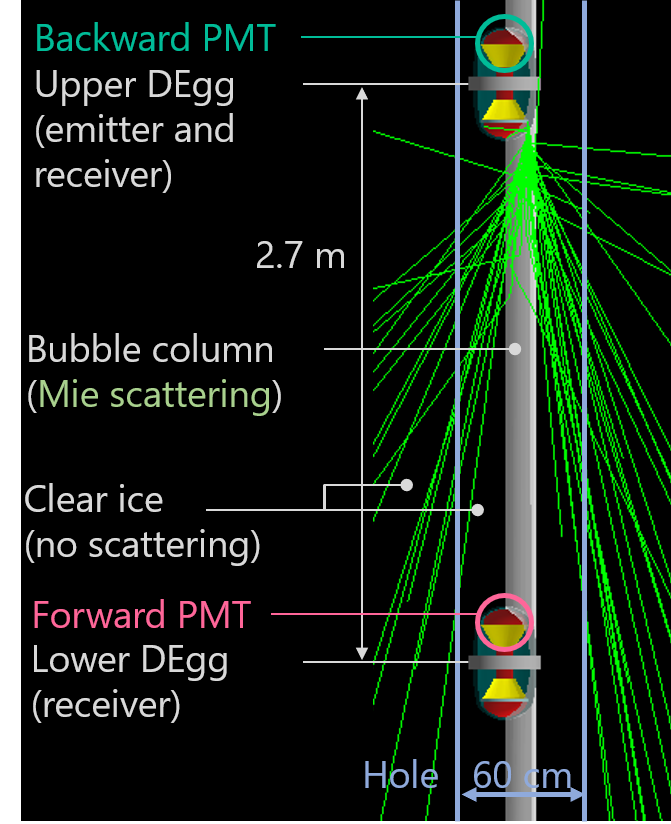}\label{fig:sim}}\\
  \end{tabular}&
 \begin{tabular}{c}
  \subfigure[]{\includegraphics[width=.35\linewidth]{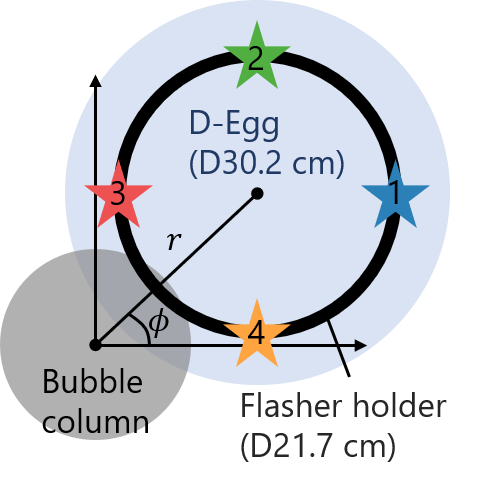}\label{fig:rphi}}\\
\subfigure[]{\includegraphics[width=.21\linewidth]{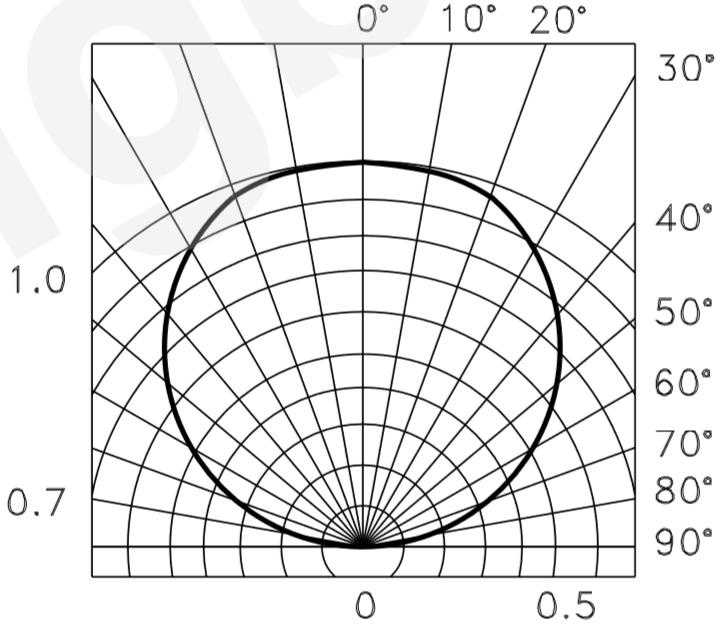}\label{fig:LEDangle}}
 \end{tabular}
\end{tabular}
\caption{a) Depiction of the simulation: two D-Egg modules are located (partially) inside the bubble column. Photons from the LED originating from the upper D-Egg module are Mie scattered in the bubble column. b) Position parameters $r$ and $\phi$. The stars represent the four downward-facing LEDs. $\phi=0$ when LED1 is the furthest from the center. c) Angular distribution of the LED with a viewing angle of $120^\circ$ \cite{ref:LED}.}
\end{figure}

Two D-Egg modules are simulated perfectly aligned in the 60\,cm diameter cylinder hole region with a vertical space of 2.7\,m.
The relative position of the D-Eggs to the bubble column is characterized by the distance $r$ from the bubble column center to the center of the D-Egg module and the rotation around the bubble column center $\phi$ (Figure~\ref{fig:rphi}).
Each of the four downward-pointing LEDs are modeled as a point light source at a wavelength of 405\,nm and a viewing angle of $120^\circ$ (Figure~\ref{fig:LEDangle}).
When the downward-pointing LEDs flash, two different PMTs can detect the photons.
The upward-facing PMT in the \textit{lower} D-Egg (forward PMT) and the upward-facing PMT in the \textit{upper} D-Egg (backward PMT) both receive these photons but due to different scattering processes (Figure~\ref{fig:sim}).

Multiple different setups were simulated, with regards to the bubble column effective scattering length, the size of the bubble column and the relative position of the D-Egg modules. This was done by changing $\lambda_\mathrm{s}$, $D$, $r$ and $\phi$ over a large range of values, and then recording the number of photoelectrons (NPE) measured by the PMTs, to eventually extract the hole ice parameters. Note, these simulations have incorporated a realistic PMT response by including the quantum efficiency and collection efficiency as measured in the laboratory \cite{ref:degg_PMT}.

\begin{figure}
\begin{minipage}[b]{0.48\linewidth}
\centering
\stackunder{\includegraphics[width=.98\linewidth]{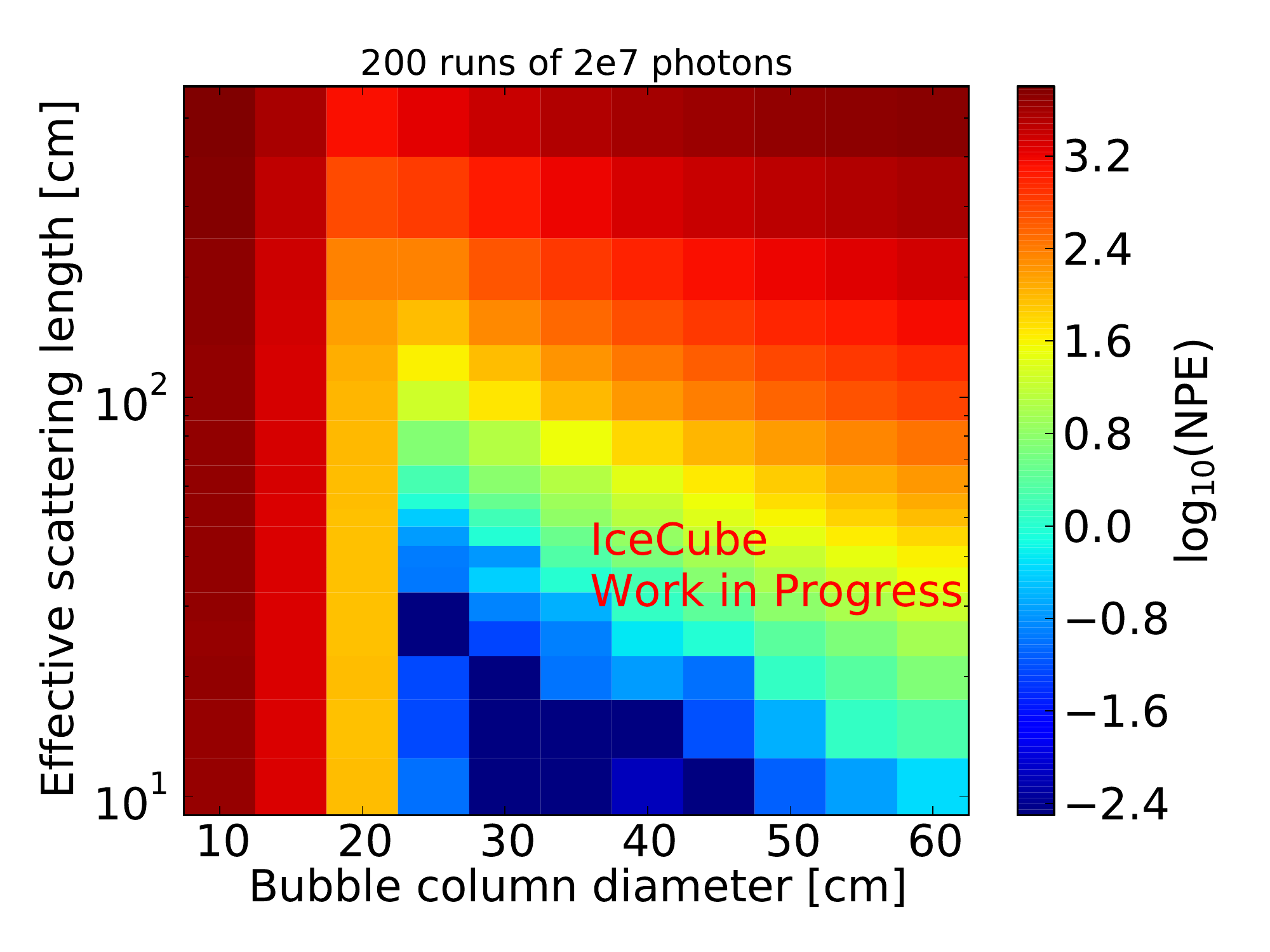}}{NPE at the forward PMT}
\end{minipage}
\begin{minipage}[b]{0.48\linewidth}
\centering
\stackunder{\includegraphics[width=.98\linewidth]{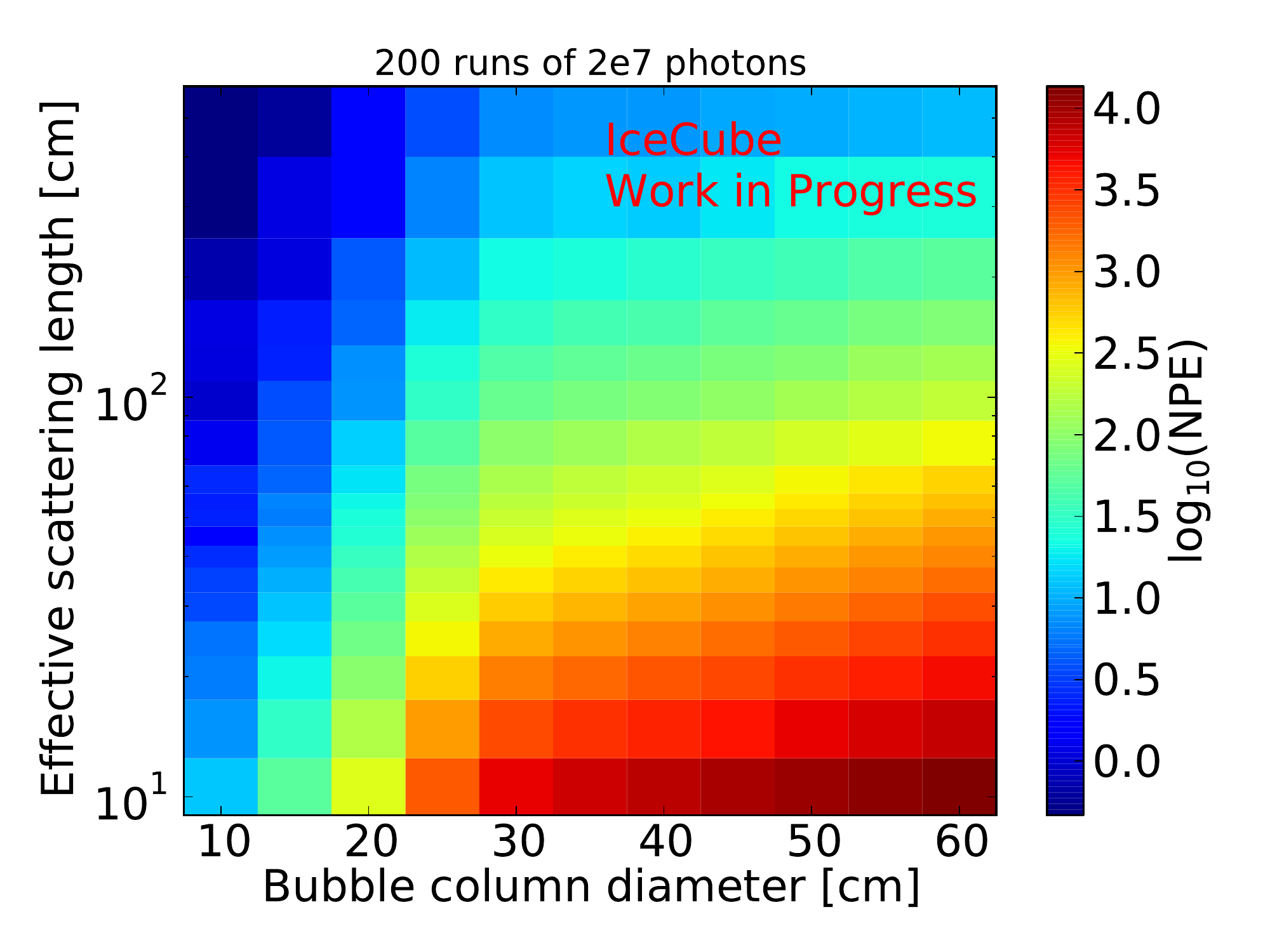}}{NPE at the backward PMT}
 \end{minipage}
\caption{Simulation results when $r=0$, for NPE averaged over 200 runs.}\label{fig:r0}
\end{figure}

Figure~\ref{fig:r0} shows the results where the D-Eggs are at the center of the bubble column ($r=0$), for the forward and backward PMT, respectively. 
The backward PMT detects more photoelectrons (PEs) when the scattering length becomes shorter as well as the case when the bubble column becomes larger. This is because the PMT observes photons which have suffered scattering in the bubble column.
Therefore, the increased scattering of photons reduces the rate observed at the forward PMT.
In cases when the bubble column is smaller than around 20\,cm, the majority of photons hit the PMT directly, avoiding the bubble column entirely.
However if $D > 20$\,cm, it covers the LED, trapping the photons, resulting in increased detection of PEs at the forward PMT.

\begin{figure}
\begin{minipage}[b]{0.48\linewidth}
\centering
\stackunder{\includegraphics[width=.98\linewidth]{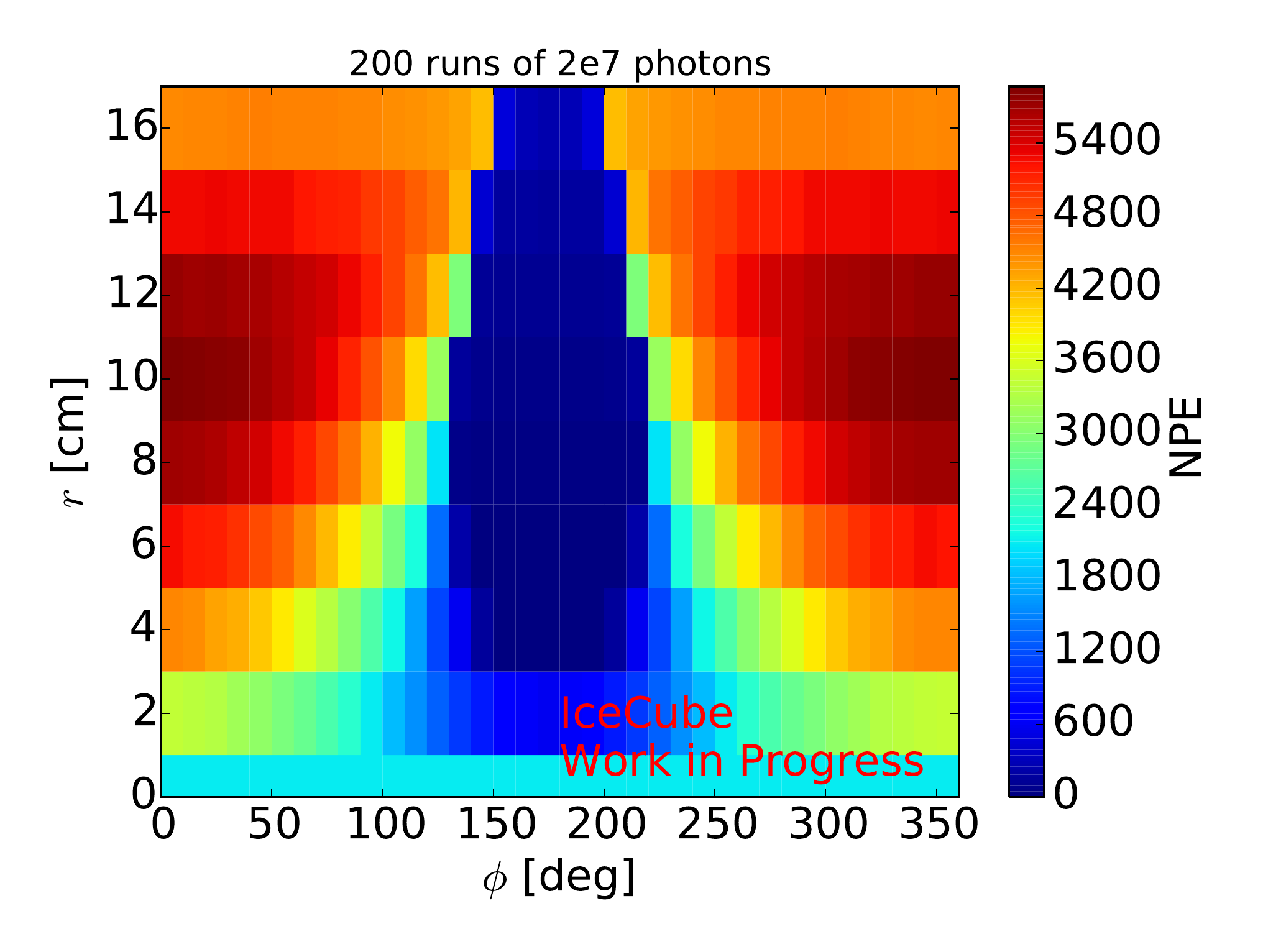}}{NPE at the forward PMT}
\end{minipage}
\begin{minipage}[b]{0.48\linewidth}
\centering
\stackunder{\includegraphics[width=.98\linewidth]{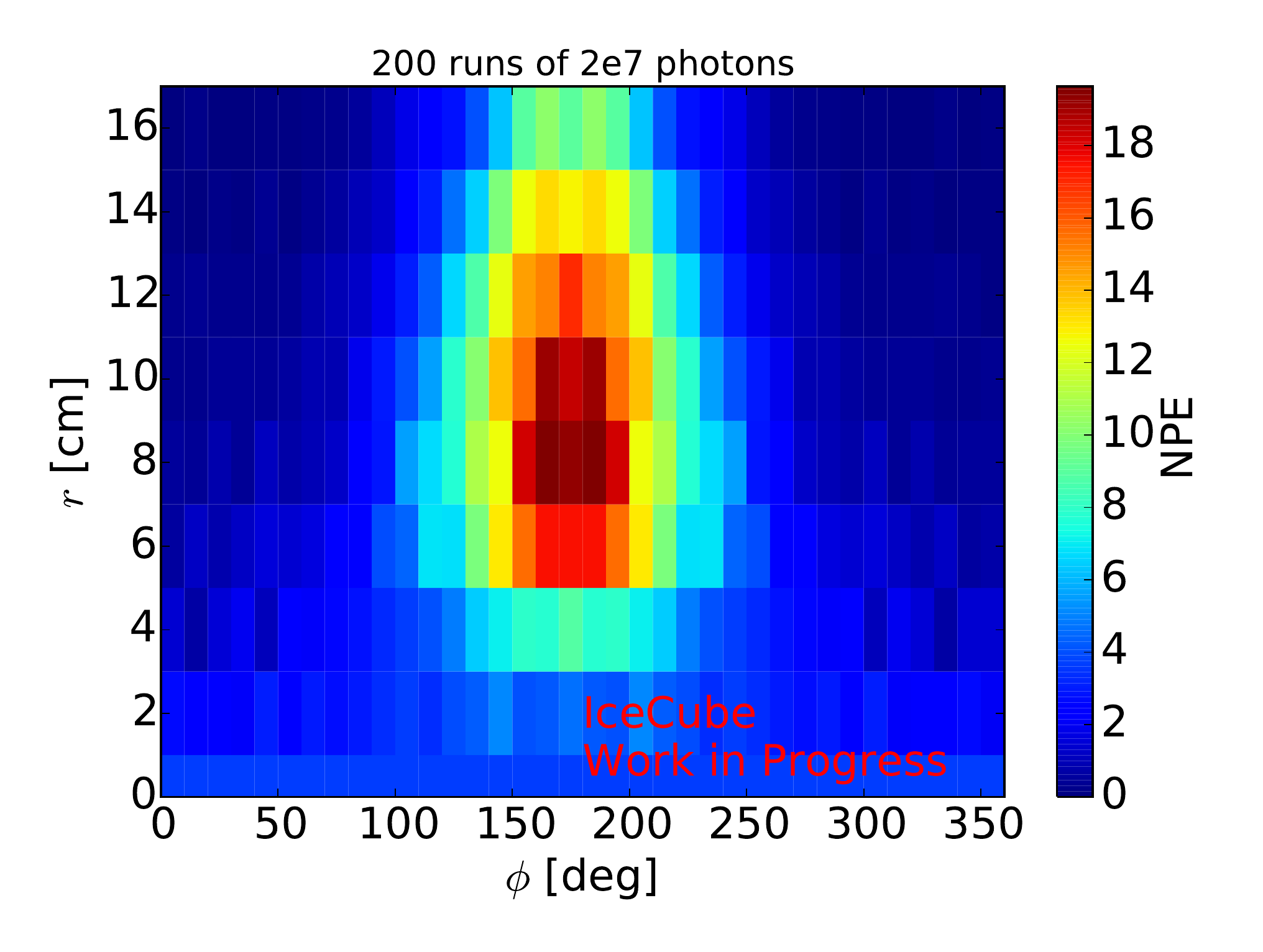}}{NPE at the backward PMT}
\end{minipage}
\caption{Simulation results when LED1 flashes, $\lambda_\mathrm{e}=100\,\mathrm{cm}$ and $D=15\,\mathrm{cm}$, for NPE averaged over 200 runs.}\label{fig:column15_rphi}
\end{figure}

\begin{figure}
\begin{minipage}[b]{0.48\linewidth}
\centering
\stackunder{\includegraphics[width=.98\linewidth]{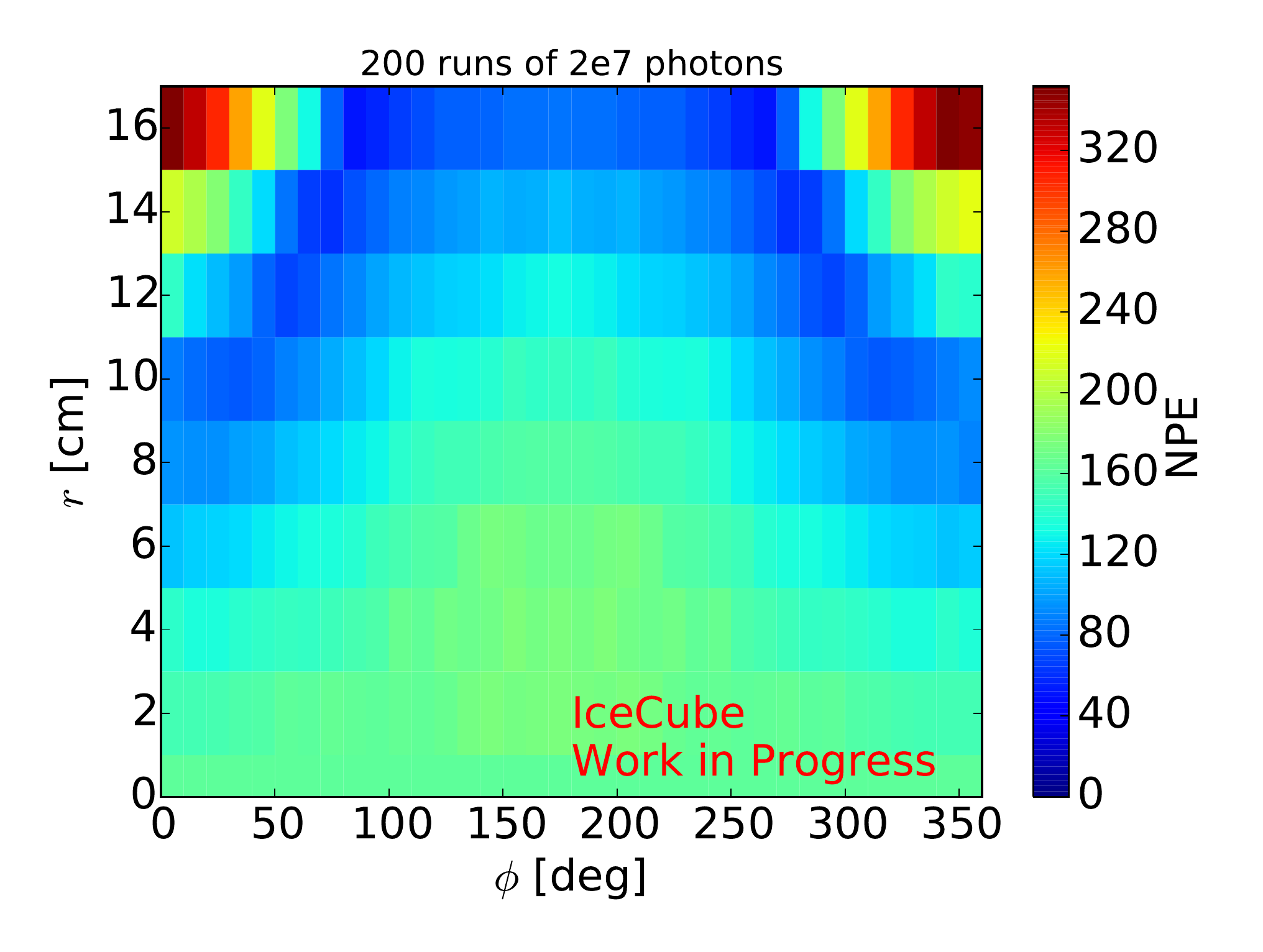}}{NPE at the forward PMT}
\end{minipage}
\begin{minipage}[b]{0.48\linewidth}
\centering
\stackunder{\includegraphics[width=.98\linewidth]{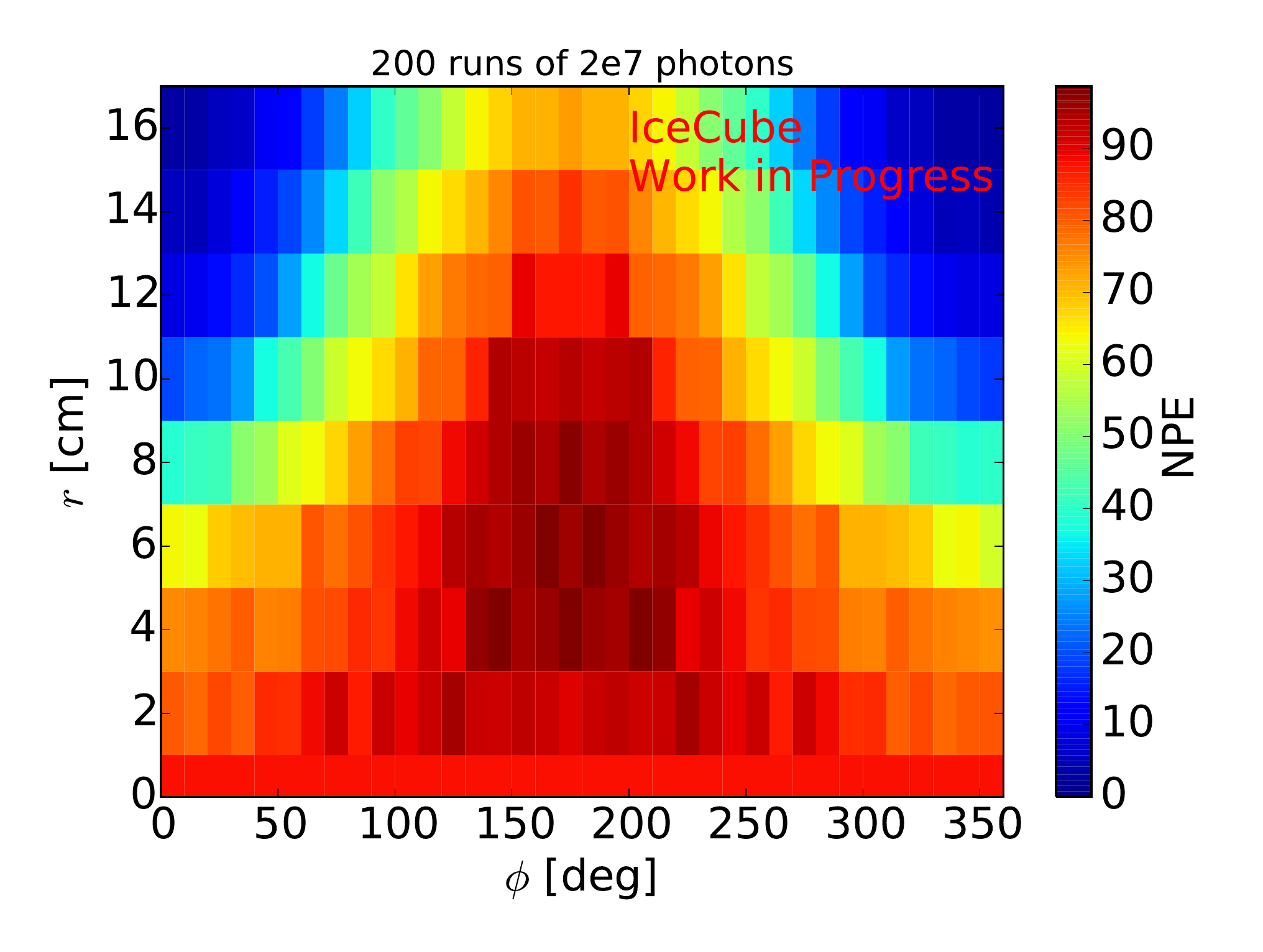}}{NPE at the backward PMT}
\end{minipage}
\caption{Simulation results when LED1 flashes, $\lambda_\mathrm{e}=100\,\mathrm{cm}$ and $D=40\,\mathrm{cm}$, for NPE averaged over 200 runs.}\label{fig:column40_rphi}
\end{figure}

Figure~\ref{fig:column15_rphi} shows the position dependence of NPE at the forward PMT and the backward PMT for a small (15\,cm) bubble column
and Figure~\ref{fig:column40_rphi} shows that for a large (40\,cm) bubble column.
In the case of the forward PMT, the NPE increases as the LED goes further from the bubble column, because of reduced photon scattering, while the backward PMT detects more PEs as the LED approaches the bubble column, because the bubble column acts as a light-guide to the PMT.

%\FloatBarrier

\section{Extracting hole ice parameters}
From the simulation, the NPE detected for both the forward and backward PMTs has been calculated over a large range of $\lambda_\mathrm{s}$, $D$, $r$ and $\phi$. However, inaccuracy in calibration of the LED intensity leads to a large uncertainty in the NPE measurements.
To minimize the effect of the variation in absolute LED intensity, the ratio of the NPE at the forward PMT and the backward PMT are constructed as an observable.
We tested a likelihood fit using the Monte Carlo described in Section \ref{monte_carlo_simulation}, in order to understand how precisely the hole ice parameters can be estimated.
To construct the ratio histograms the following assumptions are used:
\begin{itemize}
    \item The range of LED intensities follows a Gaussian distribution with a standard deviation of 30\%.
    \vspace{-0.2cm}
    \item NPE follows a Poisson distribution with the mean NPE determined from simulations as a function of the LED intensity.
    \vspace{-0.2cm}
    \item The charge resolution is characterized by a Gaussian distribution with a standard deviation of 20\%.
\end{itemize}

To perform a likelihood fit, the binned Poisson likelihood $\mathcal{L}$ for test data $\bm{n}$ and a given set of hole ice parameters $\bm{\theta}$ is constructed as:
\begin{eqnarray}
\mathcal{L}\left(\bm{n}|\bm{\nu}\left(\bm{\theta}\right)\right)=\prod_{l=1}^4 \prod_{i=1}^{N_\mathrm{bin}}\frac{{{\nu_i^l}\left(\bm{\theta}\right)}^{n_i^l}\mathrm{e}^{-\nu_i^l\left(\bm{\theta}\right)}}{n_i^l!}~,
\end{eqnarray}
where $l$ is the LED number, and $N_\mathrm{bin}\left(=10\right)$ is the number of  bins in a ratio histogram. $i$ is the bin number, $\nu_i$ is the number of events predicted to fill bin $i$, and $n_i$ is the number of events from the test data that filled bin $i$. $\bm{\theta}$ is a vector composed of the four hole ice parameters being fitted: $\lambda_\mathrm{e}$, $D$, $r$ and $\phi$.

\begin{figure}
\begin{minipage}[b]{0.48\linewidth}
\centering
\stackunder{\includegraphics[width=.94\linewidth]{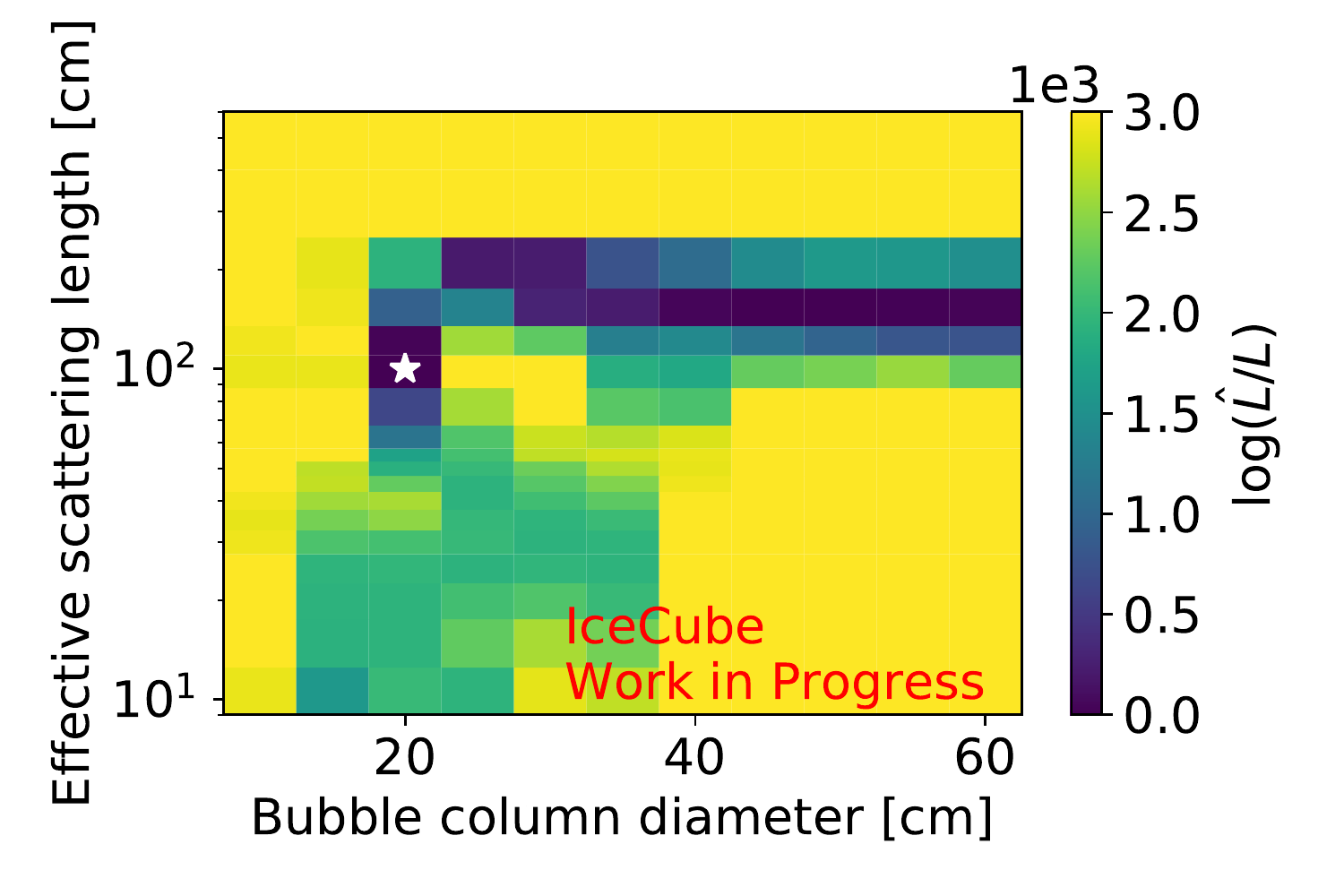}}{${\lambda_\mathrm{e}}_\mathrm{true}=100\,\mathrm{cm}, D_\mathrm{true}=20\,\mathrm{cm}$, $r_\mathrm{true}=0$}
\end{minipage}
\begin{minipage}[b]{0.48\linewidth}
\centering
\stackunder{\includegraphics[width=.94\linewidth]{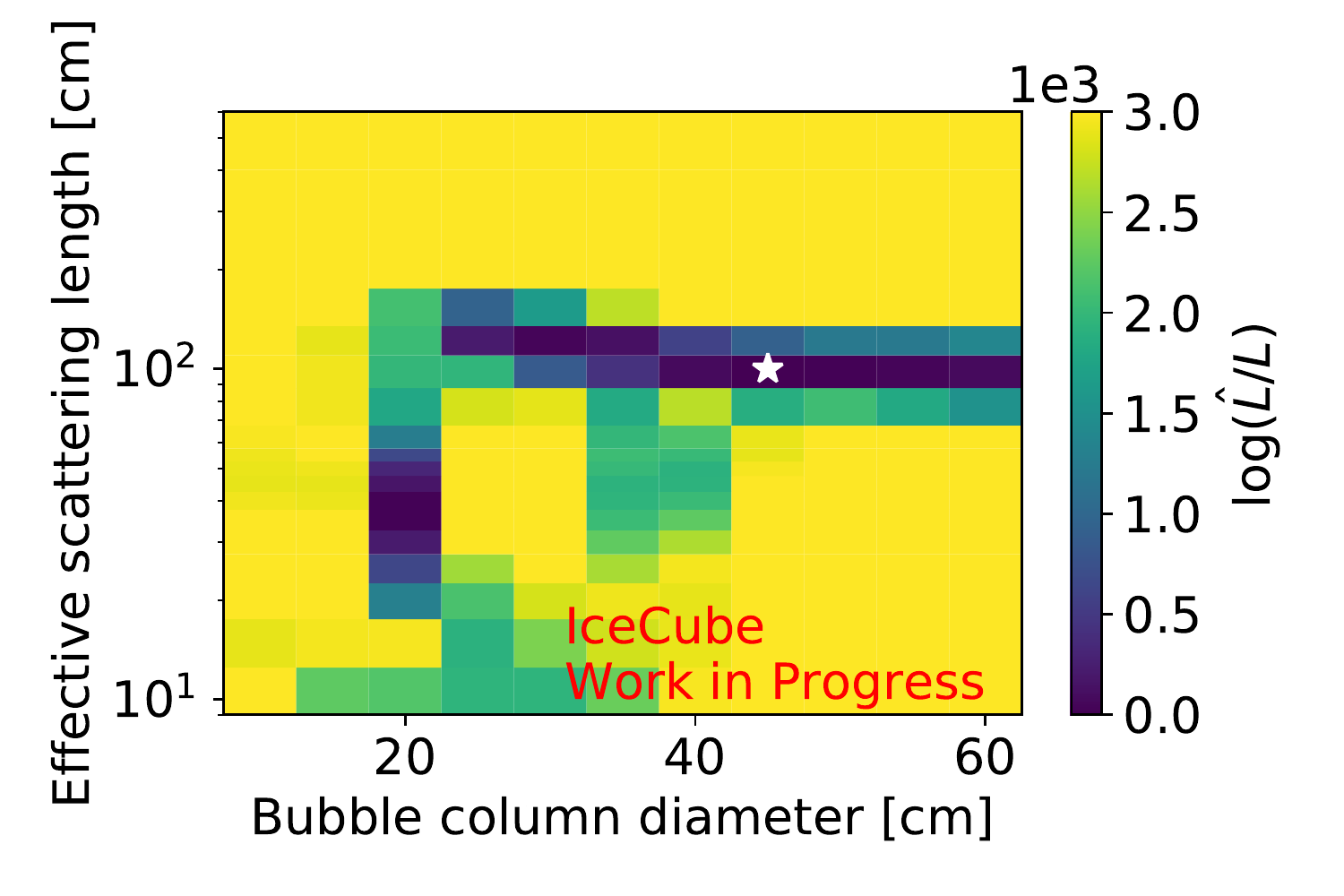}}{${\lambda_\mathrm{e}}_\mathrm{true}=100\,\mathrm{cm}, D_\mathrm{true}=45\,\mathrm{cm}$, $r_\mathrm{true}=0$}
\end{minipage}\\
\begin{minipage}[b]{0.48\linewidth}
\centering
\stackunder{\includegraphics[width=.94\linewidth]{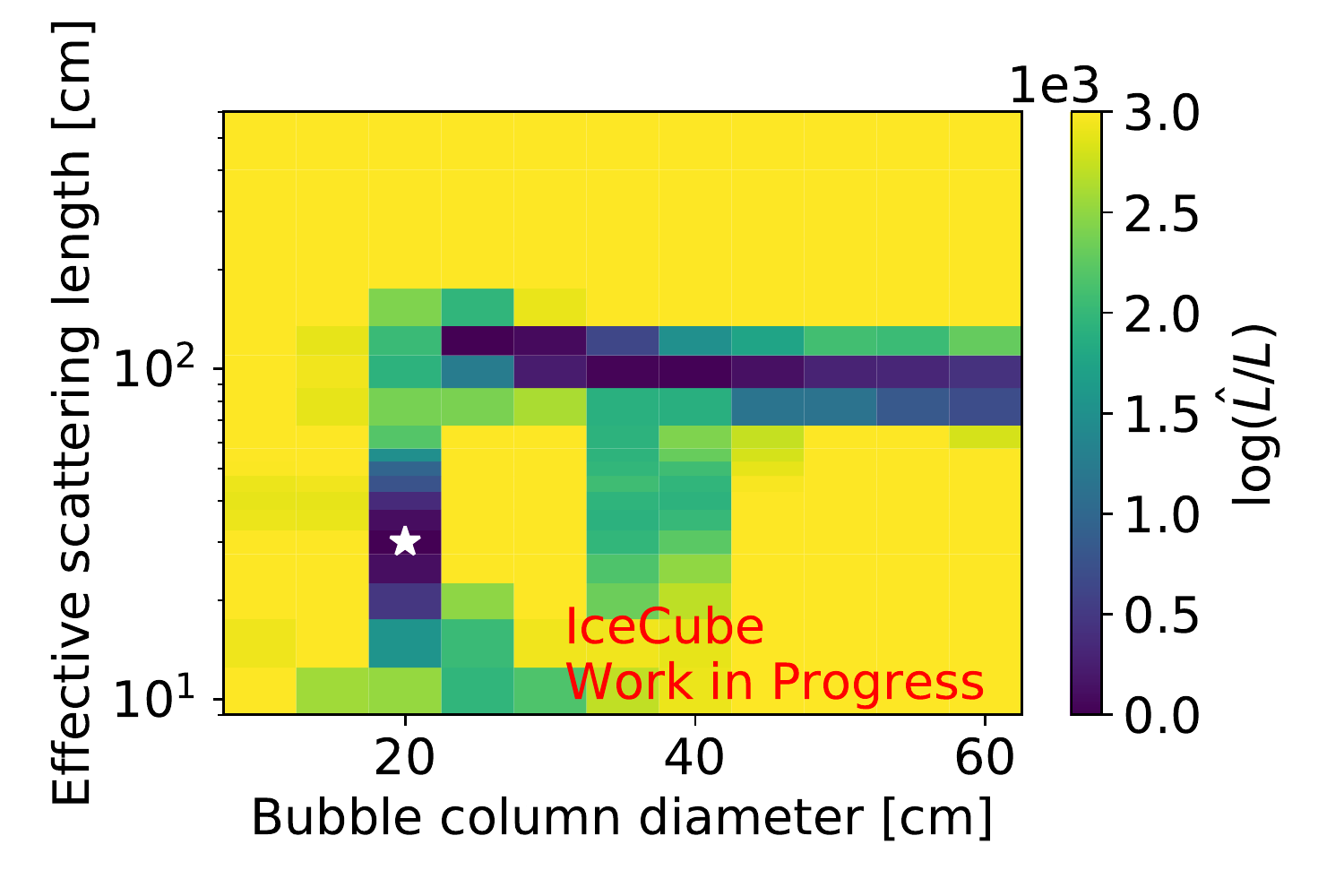}}{${\lambda_\mathrm{e}}_\mathrm{true}=30\,\mathrm{cm}, D_\mathrm{true}=20\,\mathrm{cm}$, $r_\mathrm{true}=0$}
\end{minipage}
\begin{minipage}[b]{0.48\linewidth}
\centering
\stackunder{\includegraphics[width=.94\linewidth]{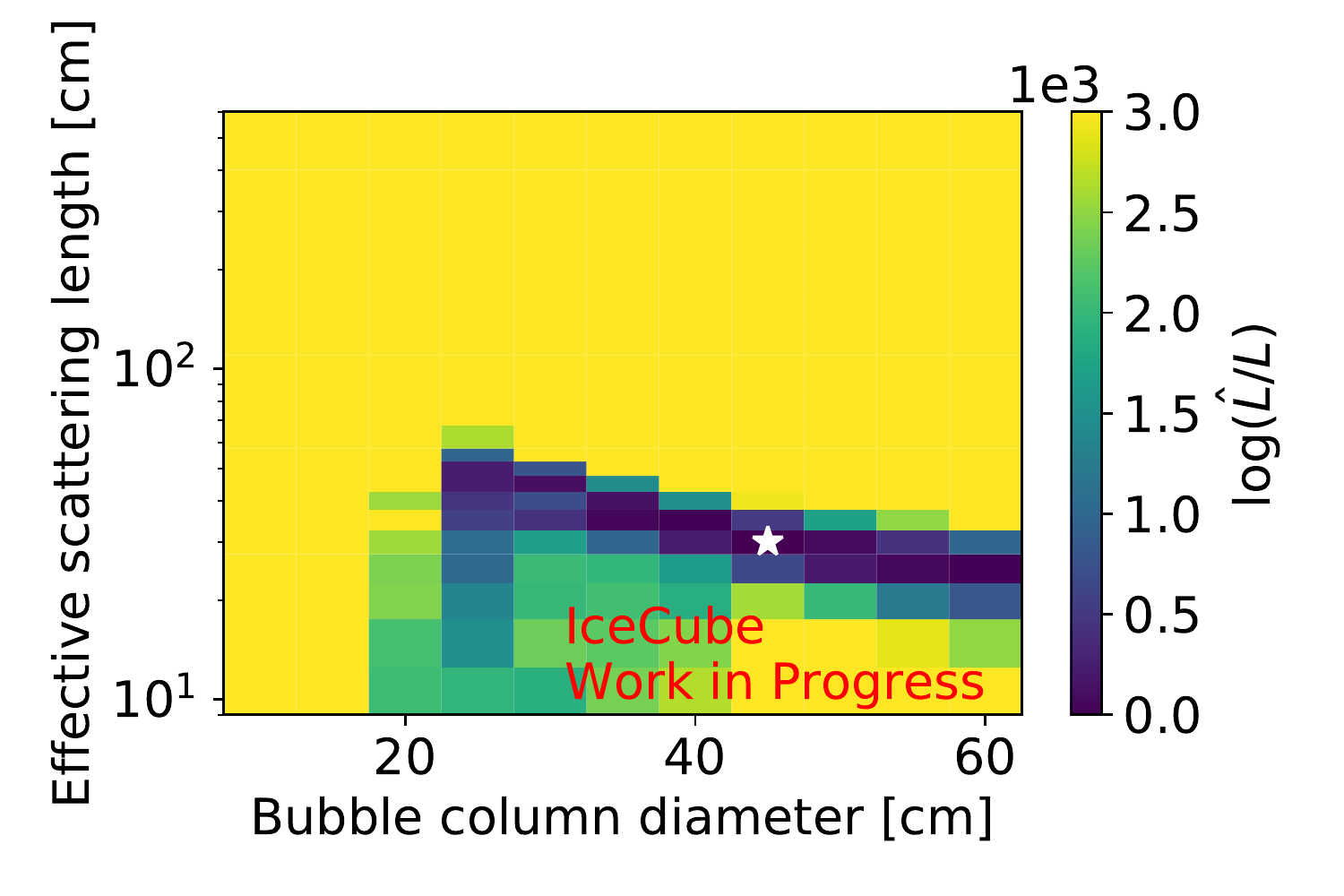}}{${\lambda_\mathrm{e}}_\mathrm{true}=30\,\mathrm{cm}, D_\mathrm{true}=45\,\mathrm{cm}$, $r_\mathrm{true}=0$}
\end{minipage}\\
\begin{minipage}[b]{0.48\linewidth}
\centering
\stackunder{\includegraphics[width=.94\linewidth]{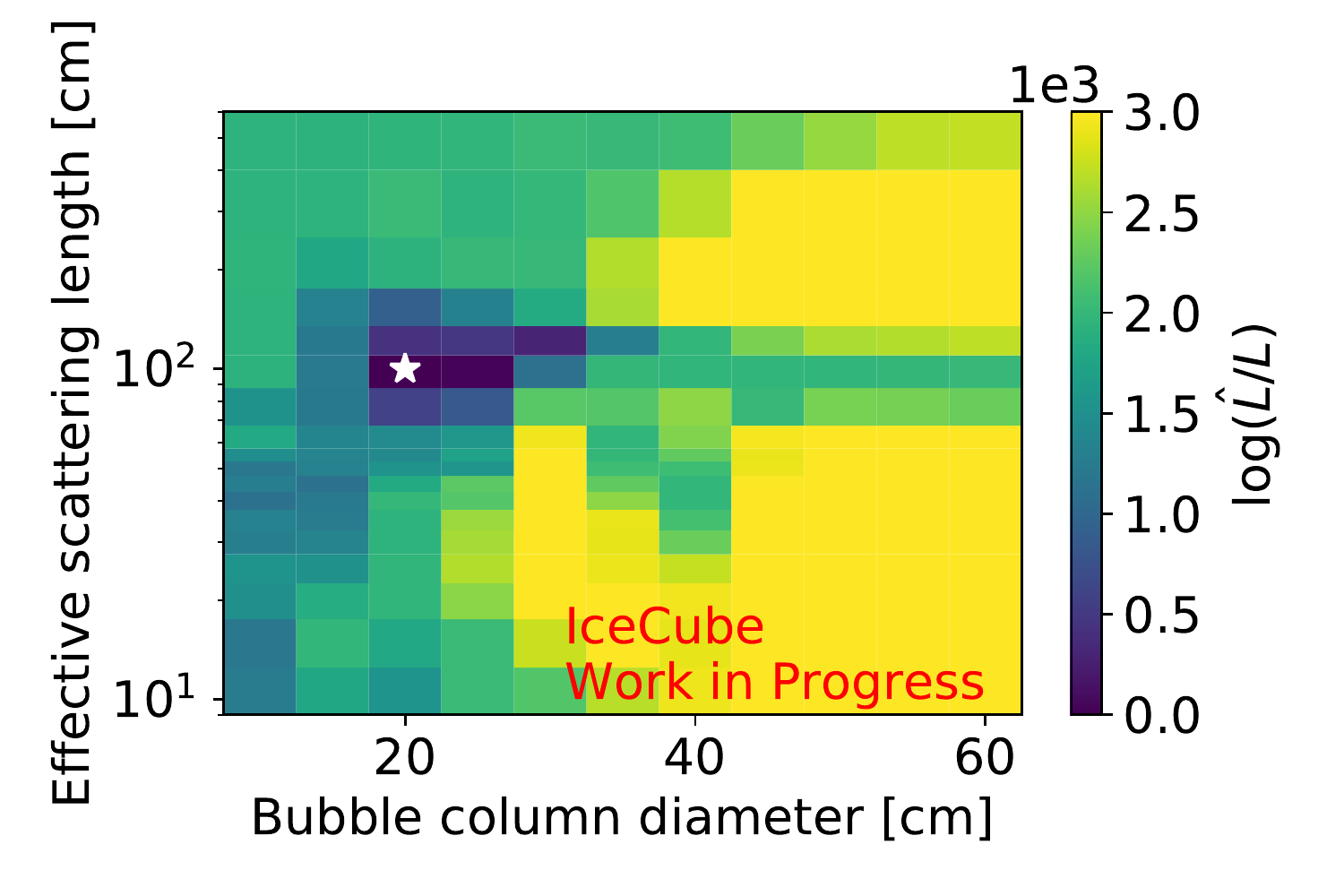}}{\begin{tabular}{c}${\lambda_\mathrm{e}}_\mathrm{true}=100\,\mathrm{cm}, D_\mathrm{true}=20\,\mathrm{cm}$,\\$r_\mathrm{true}=10\,\mathrm{cm}, \phi_\mathrm{true}=45^\circ$\end{tabular}}
\end{minipage}
\begin{minipage}[b]{0.48\linewidth}
\centering
\stackunder{\includegraphics[width=.94\linewidth]{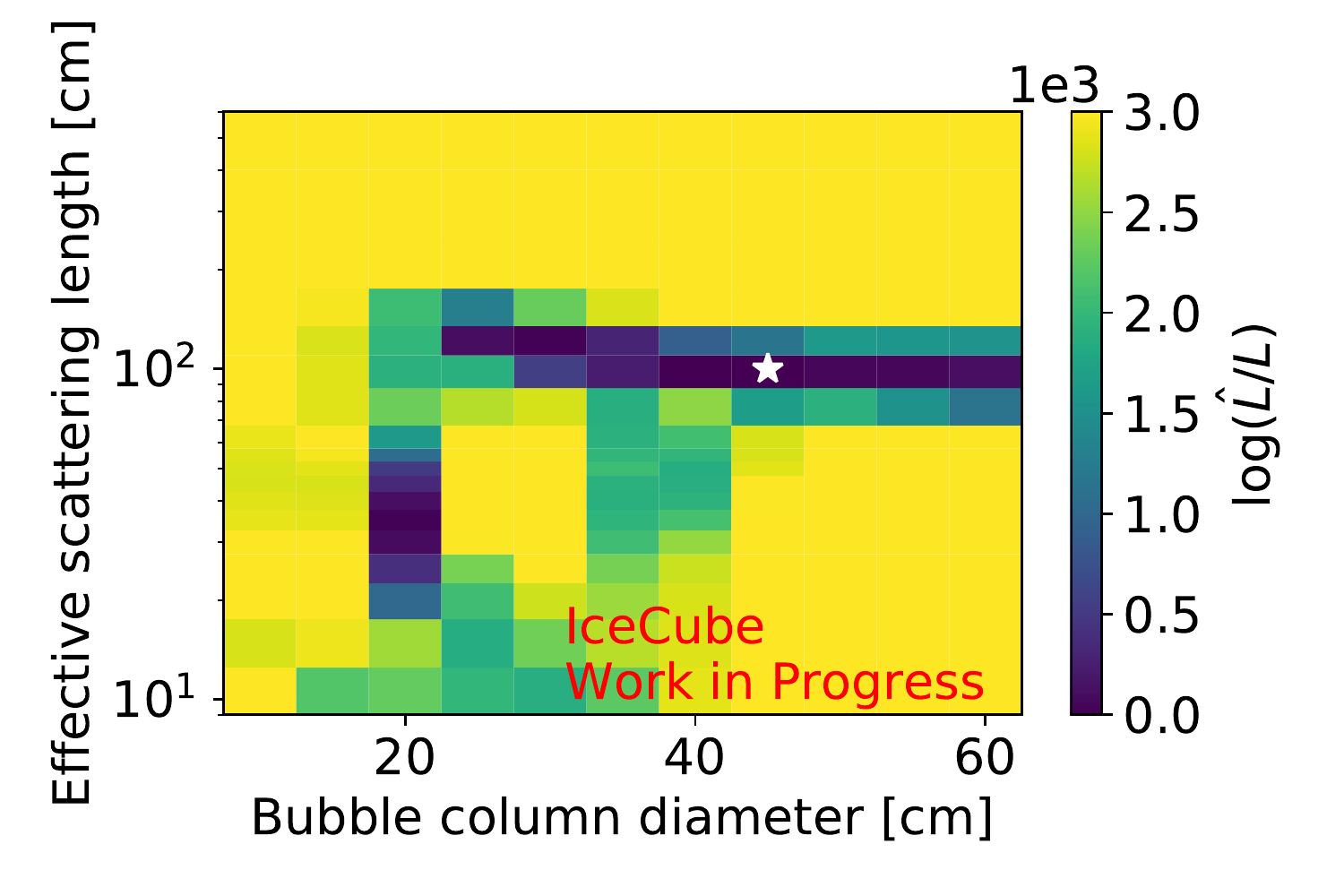}}{\begin{tabular}{c}${\lambda_\mathrm{e}}_\mathrm{true}=100\,\mathrm{cm}, D_\mathrm{true}=45\,\mathrm{cm}$,\\$r_\mathrm{true}=10\,\mathrm{cm}, \phi_\mathrm{true}=45^\circ$\end{tabular}}
\end{minipage}\\
\begin{minipage}[b]{0.48\linewidth}
\centering
\stackunder{\includegraphics[width=.94\linewidth]{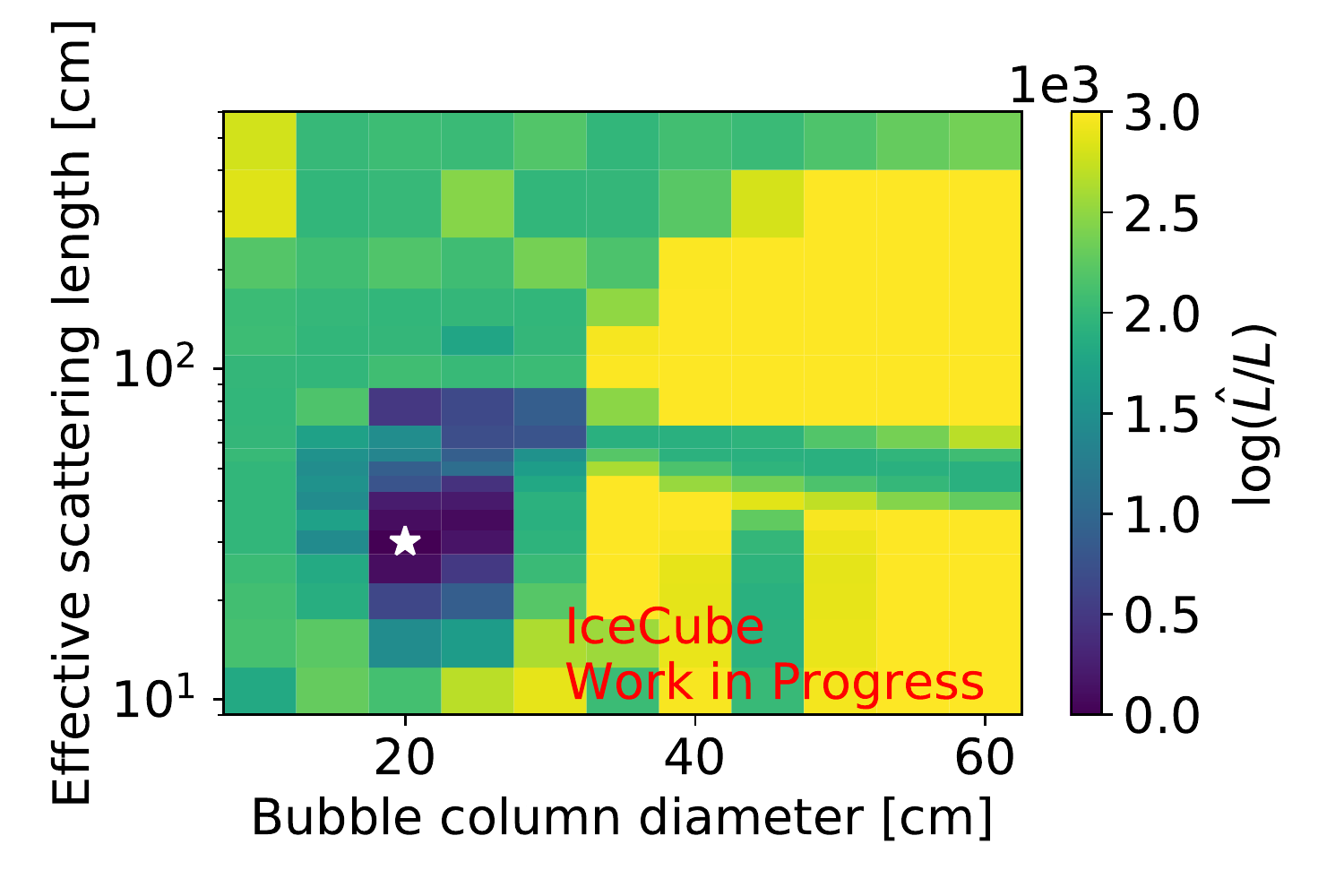}}{\begin{tabular}{c}${\lambda_\mathrm{e}}_\mathrm{true}=30\,\mathrm{cm}, D_\mathrm{true}=20\,\mathrm{cm}$, \\$r_\mathrm{true}=10\,\mathrm{cm}, \phi_\mathrm{true}=45^\circ$\end{tabular}}
\end{minipage}
\begin{minipage}[b]{0.48\linewidth}
\centering
\stackunder{\includegraphics[width=.94\linewidth]{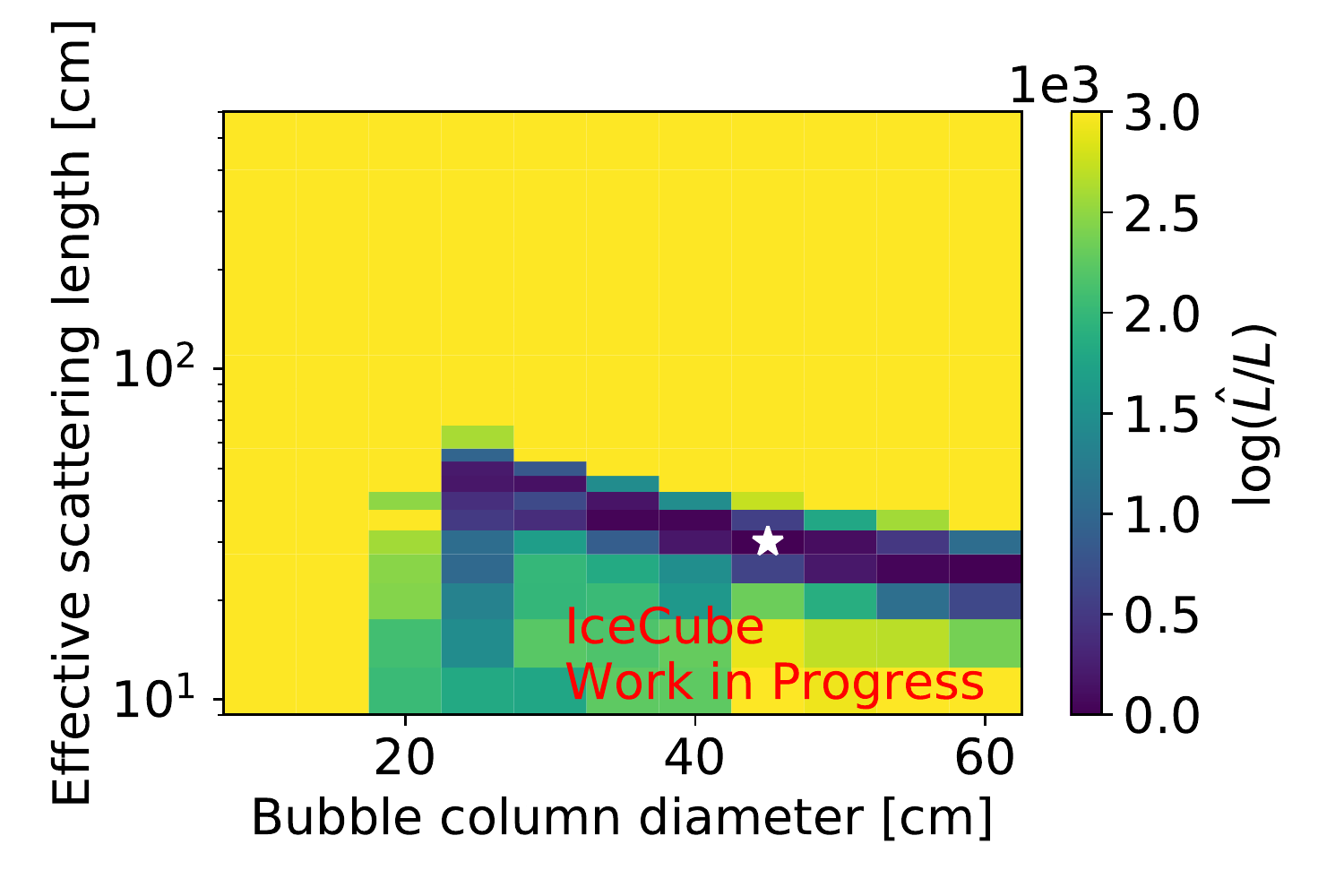}}{\begin{tabular}{c}${\lambda_\mathrm{e}}_\mathrm{true}=30\,\mathrm{cm}, D_\mathrm{true}=45\,\mathrm{cm}$, \\$r_\mathrm{true}=10\,\mathrm{cm}, \phi_\mathrm{true}=45^\circ$\end{tabular}}
\end{minipage}
\caption{Log likelihood ratio for each MC truth. Best fit parameter bin is marked with a white star.}\label{fig:lnLratio}
\end{figure}

Figure~\ref{fig:lnLratio} shows the log likelihood ratio for various  hole ice parameters, where $r$ and $\phi$ are treated as nuisance parameters.
The best fit point is always found in the same bin as the MC parameters, and the size of the 3$\sigma$ region is smaller than the current bin size.
These results indicate that a measurement of the hole ice with the D-Egg modules using the downward-facing LEDs has the potential to determine the effective scattering length in the bubble column to within $\pm20$\% and the bubble column diameter to within $\pm5\,\mathrm{cm}$.
These results are limited by the chosen binning scheme, and future studies will benefit from finer sampling (decreased bin size).
Currently, these results do not include systematic errors due to bulk ice scattering, potential uncertainty in the D-Egg alignment, and obstruction of the signal due to cabling.

%\newpage
%\FloatBarrier

\section{Summary and outlook}
We have presented a Monte Carlo-based sensitivity study of hole ice measurements using the D-Egg optical modules to be deployed in the IceCube Upgrade.
The D-Egg modules are able to measure the refrozen ice properties, including those of the bubble column, by utilizing four downward-facing flasher LEDs, which illuminate the ice around the strings of D-Egg modules.
By simulating the geometry of the in-ice system and varying parameters of the hole ice, expectations of the observed NPE were calculated for both forward and backward facing PMTs.
Using the ratio of NPE at the two PMTs, a likelihood fit was performed where the Monte Carlo truth parameter bin is always found as the most likely point.
Critically, these results take into account realistic properties of the PMTs (charge resolution, quantum efficiency, collection efficiency) and minimize the impact of uncertainty related to the LED intensity.
That will enable us to determine the hole ice properties such as the effective scattering length in the bubble column to within $\pm20$\% and the bubble column diameter to within $\pm5\,\mathrm{cm}$.
This simulation and these results form a critical baseline for future in-situ measurements using the D-Egg modules and the downward-facing LEDs.

\bibliographystyle{icrc}
\bibliography{references}
%\printbibliography

\end{document}